\documentstyle[epsfig,psfig]{texas}

\begin{document}

\title{OPTICAL STUDIES OF THE\\ X-RAY TRANSIENT XTE J2123--058 -- II.\\
PHASE RESOLVED SPECTROSCOPY}
\author{Robert I. HYNES$^{1}$\footnote{Present address: Physics
department, The Open University, Milton Keynes, MK7 6AA, UK}, Philip A. CHARLES$^{2}$,\addtocounter{footnote}{-1} 
Carole A. HASWELL$^{1}\footnotemark$, Jorge CASARES$^{3}$, Cristina ZURITA$^{3}$}
\address{(1) Astronomy Centre, University of Sussex\\
Falmer, Brighton, BN1 9QJ, UK}
\address{(2) Astrophysics, University of Oxford\\
Nuclear and Astrophysics Laboratory, Keble Road, Oxford, OX1 3RH, UK}
\address{(3) Instituto de Astrofisica de Canarias,\\ 38200 La Laguna,
       Tenerife, Spain}
{\rm Email: R.I.Hynes@open.ac.uk}

\begin{abstract}
We present time-resolved spectroscopy of the soft X-ray transient
XTE~J2123--058 in outburst.  Spectral coverage of 3700--6700\,\AA\ was
achieved spanning two orbits of the binary.  The strongest emission
lines are He\,\textsc{ii} 4686\,\AA\ and C\,\textsc{iii} /N\,\textsc{iii}
4640\,\AA\ (Bowen blend).  Other weak emission lines of He\,\textsc{ii} and
C\,\textsc{iv} are present and Balmer lines show a complex structure,
possibly contaminated by He\,\textsc{ii}.  He\,\textsc{ii} 4686\,\AA\ and
C\,\textsc{iii}/N\,\textsc{iii} 4640\,\AA\ show different orbital light
curves indicating an origin in different regions.  He\,\textsc{ii}
4686\,\AA\ profiles show a complex multiple S-wave structure.  Doppler
tomography reveals this emission is not associated with the companion
star, and occurs at velocities too low for Keplerian disk material.
It can possibly be associated with overflowing or splashing stream
material.  The optical spectrum approximates a steep blue power-law,
consistent with emission on the Rayleigh-Jeans tail of a black body
spectrum.  Orbital modulations show no wavelength dependence; this is
as expected if both disk and companion star are hot enough for the
peak of their spectral energy distributions to be in the UV.  The hot
continuum and presence of high-excitation emission lines indicate
strong X-ray heating.

\end{abstract}

\section{Introduction}

The X-ray transient XTE~J2123--058 was discovered by the {\it RXTE}
satellite on 1998 June 27 (Levine, Swank and Smith 1998, \cite{L98}).
An optical counterpart was promptly identified by
Tomsick et al.\ (1998a, \cite{T98a}).  The discovery of apparent
Type-I X-ray bursts (Takeshima and Strohmayer 1998, \cite{TS98})
indicated that the compact object was a neutron star.  Interest in the
object increased dramatically when Casares et al.\ (1998, \cite{C98})
reported the presence of a strong optical modulation and attributed
this to an eclipse; the orbital period was subsequently determined to
be 6.0-hr both photometrically (Tomsick et al.\ 1998b, \cite{T98b};
Ilovaisky \& Chevalier 1998, \cite{IC98}) and spectroscopically (Hynes
et al.\ 1998, \cite{H98}).  Tomsick et al.\ (1998b, \cite{T98b})
suggested that the 0.9-mag modulation is likely actually due to the
changing aspect of the heated companion in a high inclination system,
although partial eclipses appear also to be superposed on this
(Zurita, Casares \& Hynes 1998, \cite{ZCH98}).  In this paper we
present the results of our spectrophotometric study of XTE~J2123--058
using the William Herschel Telescope (WHT), La Palma.  Our photometric
observations are described in a companion paper in this proceedings,
Zurita et al.\ (1999, hereafter Paper I, \cite{Z99}).

\section{Our dataset}

We observed XTE J2123--058 through two 6-hr binary orbits on 1998 July
19--20.  We used the blue arm of the ISIS dual-beam spectrograph on
the 4.2-m William Herschel Telescope to obtain 28 spectra.  The R300B
grating combined with an EEV $4096\times2048$ CCD gave an unvignetted
coverage of $\sim$4000--6500\,\AA\ with some useful data outside this
range.  An 0.7--1.0\,arcsec slit gave a spectral resolution
2.9--4.1\,\AA.  Each spectrum was calibrated relative to a second star
on the slit.  Absolute calibration was tied to the spectrophotometric
standard Feige 110 (Oke 1990, \cite{O90}).  Wavelength calibration was
obtained from a copper-argon arc lamp, with spectrograph flexure
corrected using sky emission lines.

\begin{figure}
\centering
\psfig{file=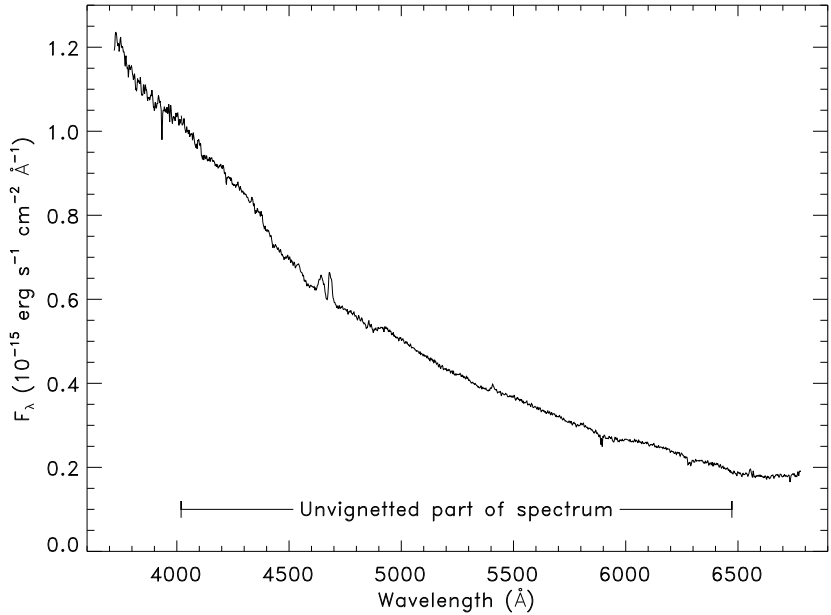,angle=90,width=12cm}
\caption{Average optical spectrum of XTE J2123--058.}
\label{SpecFig}
\end{figure}

Our average spectrum shown in Fig.\ \ref{SpecFig} is derived from a
straight sum of count rates before slit loss and extinction
corrections to maximize the signal to noise ratio.  The
spectral energy distribution was determined from an
average of calibrated spectra interpolated onto a uniform phase grid,
i.e.\ it is a uniformly weighted average over all phases.

\section{The line spectrum}

At first glance, XTE J2123--058 presents a nearly featureless blue
spectrum, with only the Bowen blend (N\,\textsc{iii}/C\,\textsc{iii}
4640\,\AA) and He\,\textsc{ii} 4686\,\AA\ prominent.  In addition,
however, a number of weaker emission lines are present, the Balmer
lines exhibit complex profiles and weak interstellar absorption
features are seen.

\begin{figure}
\centering
\noindent\psfig{file=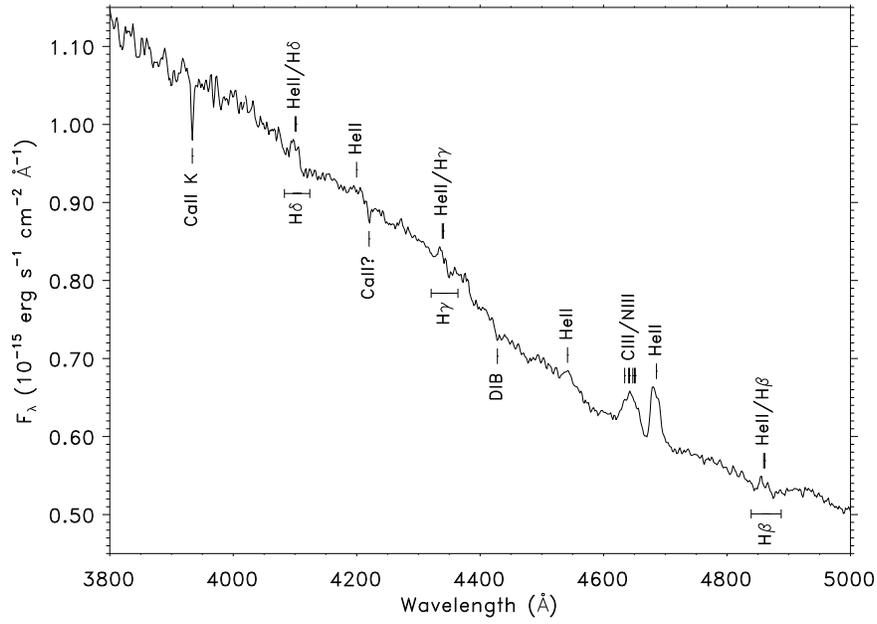,angle=90,width=12cm}
\centering
\noindent\psfig{file=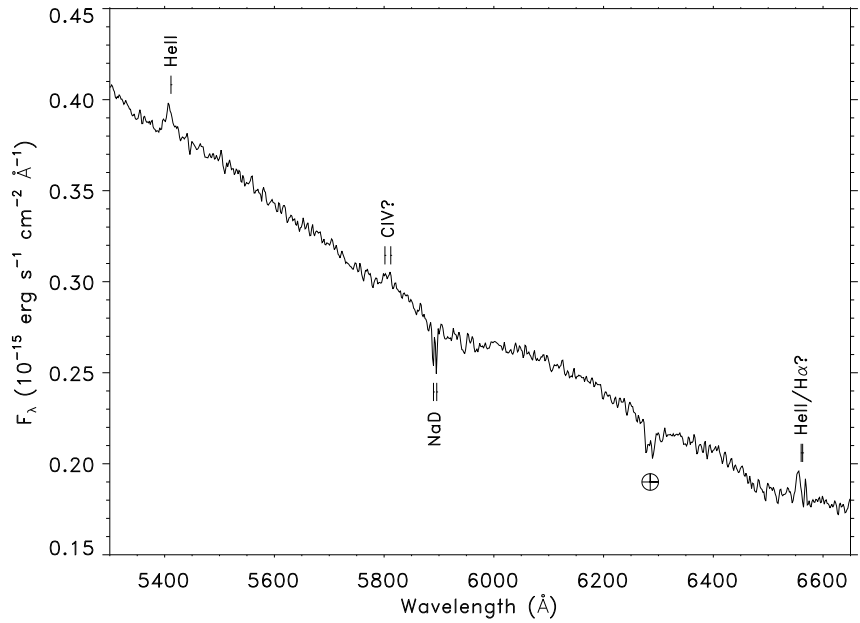,angle=90,width=12cm}
\caption{Average optical spectrum of XTE J2123--058 in more detail.  
All identified absorption or emission lines are marked.}
\label{LineSpecFig}
\end{figure}

\begin{table}[t]
\begin{center}
\begin{tabular}{rlll}
\hline
\noalign{\smallskip}
        & Identification      & Wavelength (\AA) & Comment \\
\noalign{\smallskip}
\hline
\noalign{\smallskip}
$\star$ & Ca\,\textsc{ii}\,K            & 3933.7        & 
Interstellar\\ 
        & He\,\textsc{ii}\ Br$\theta$   & 4100.0        & 
Blended with H$\delta$?\\ 
        & H$\delta$           & 4101.7        & \\
        & He\,\textsc{ii}\ Br$\eta$     & 4199.8        & \\ 
        & Ca\,\textsc{ii}\               & 4220.1        & \\ 
        & He\,\textsc{ii}\ Br$\zeta$    & 4338.7        & 
Blended with H$\gamma$?\\ 
        & H$\gamma$           & 4340.5        & \\
        & DIB                 & 4428          & 
Interstellar\\
$\star$ & He\,\textsc{ii}\ Br$\epsilon$ & 4541.6        & \\ 
$\star$ & N\,\textsc{iii}/C\,\textsc{iii} & 4640          & \\ 
$\star$ & He\,\textsc{ii}\ P$\alpha$    & 4685.7        & \\ 
        & He\,\textsc{ii}\ Br$\delta$   & 4859.3        & 
Blended with H$\beta$?\\
        & H$\beta$            & 4861.3        & \\
$\star$ & He\,\textsc{ii}\ Br$\gamma$   & 5411.5        & \\ 
        & C\,\textsc{iv}                & 5801.5,5812.1 & \\ 
$\star$ & Na\,\textsc{i}\,D             & 5890.0,5895.9 & 
Interstellar\\ 
        & He\,\textsc{ii}\ Br$\beta$    & 6560.1        & 
Blended with H$\alpha$?\\ 
        & H$\alpha$           & 6562.5        & 
\end{tabular}
\vspace{3mm}
\caption{Spectral lines detected in XTE~J2123--058; $\star$ indicates a
definite detection.}
\label{EmissionTable}
\end{center}
\end{table}

Balmer lines from H$\beta$ to H$\delta$ appear to show broad
absorption and an emission core.  The wavelength range marked
underneath each Balmer line in Fig.\ \ref{LineSpecFig} corresponds to
$\pm1500$\,km\,s$^{-1}$; this is intended to be an approximate guide
to the width, rather than a fit.  The emission core may partly be Balmer
emission but at least some is attributable to coincident
He\,\textsc{ii} lines (since we see He\,\textsc{ii} 4542\,\AA\ and
5412\,\AA\, we would expect to also see related lines such as
4859\,\AA).  This broad absorption plus narrow emission Balmer line
structure is also seen in other systems, for example the neutron star
LMXB 4U~2129+47 (Thorstensen \& Charles 1982, \cite{TC82}) and the
black hole candidate GRO~J0422+32 in outburst (Shrader et al.\ 1994,
\cite{S94}).

Based on the strength of the Na\,D1 line and the calibration of Munari
\& Zwitter (1997, \cite{MZ97}) we estimate $E(B-V) = 0.12 \pm 0.05$.

\section{Emission line behavior}

\begin{figure}
\centering
\psfig{file=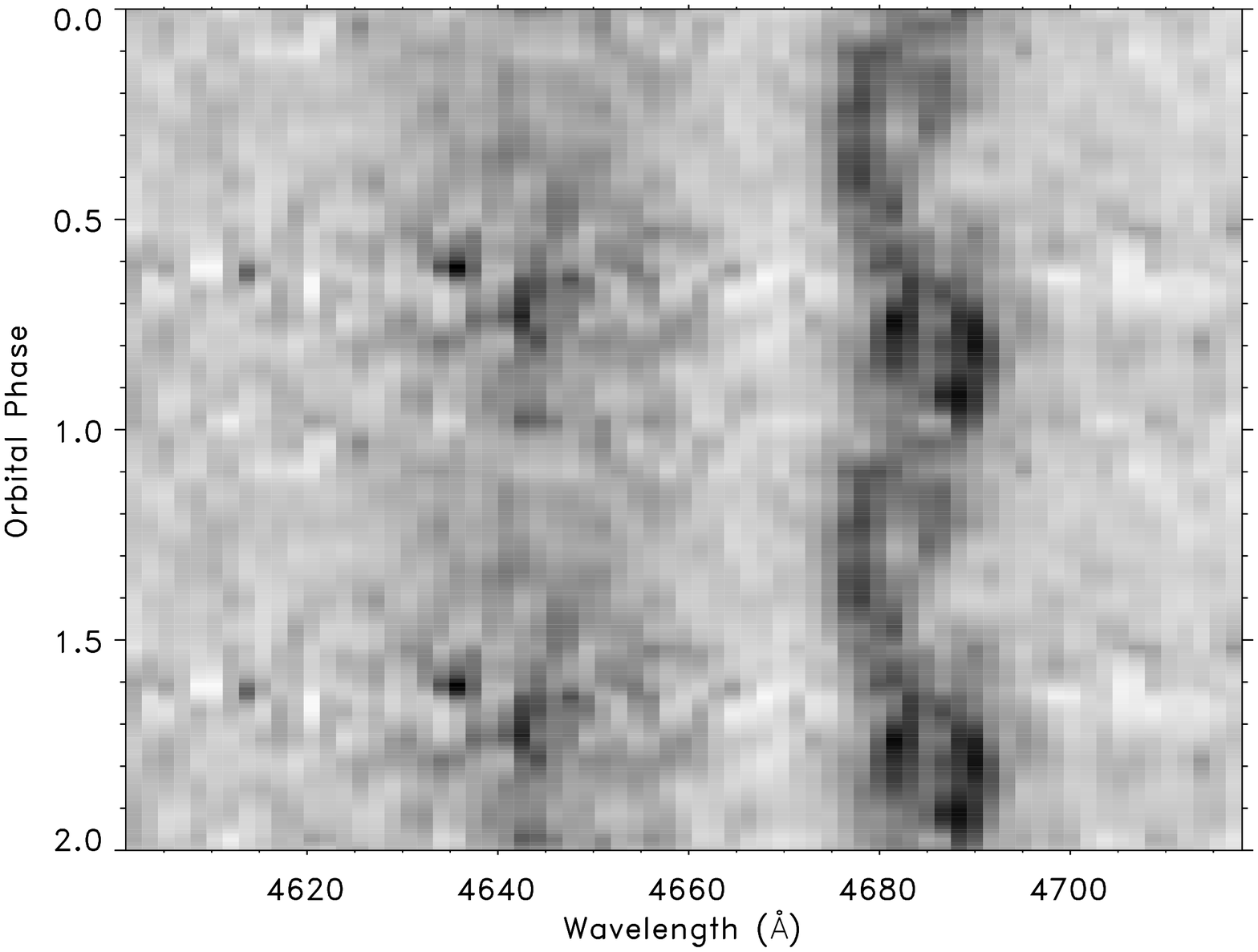,height=6.2cm}
\hspace{\fill}
\psfig{file=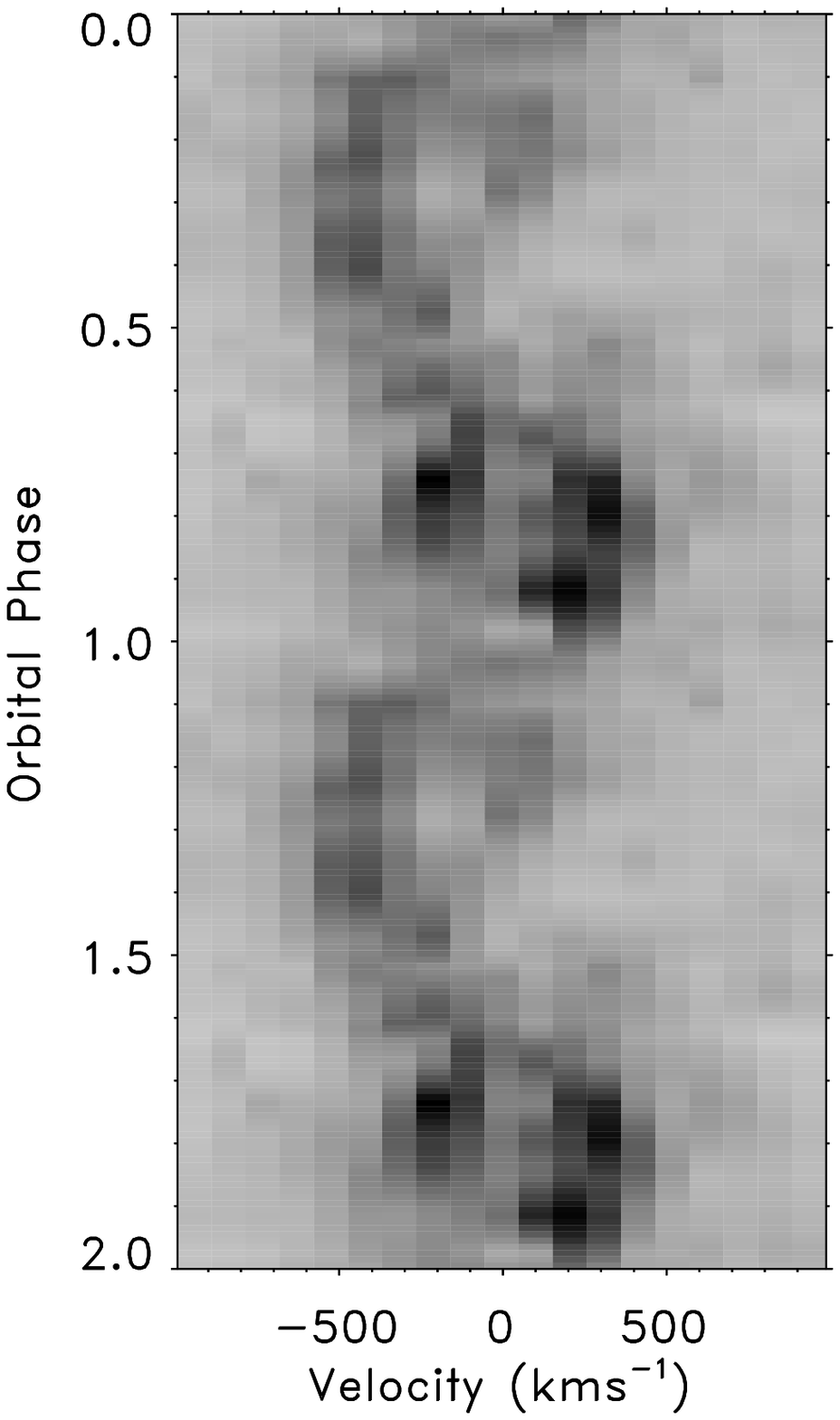,height=6.2cm}
\caption{a) Emission line trailed spectrograms of
C\,\textsc{iii}/N\,\textsc{iii} 4640\,\AA\ and He\,\textsc{ii}
4686\,\AA\ in wavelength. b) Expansion of the He\,\textsc{ii}
spectrogram in velocity.}
\label{SpectrogramFig}
\end{figure}

Both N\,\textsc{iii}/C\,\textsc{iii} 4640\,\AA\ (Bowen blend) and
He\,\textsc{ii} 4686\,\AA\ emission lines show changes in integrated flux
over an orbital cycle and He\,\textsc{ii} also reveals complex line
profile changes, with multiple S-wave components present in the
trailed spectrogram shown in Fig.\ \ref{SpectrogramFig}.

\begin{figure}
\centering
\psfig{file=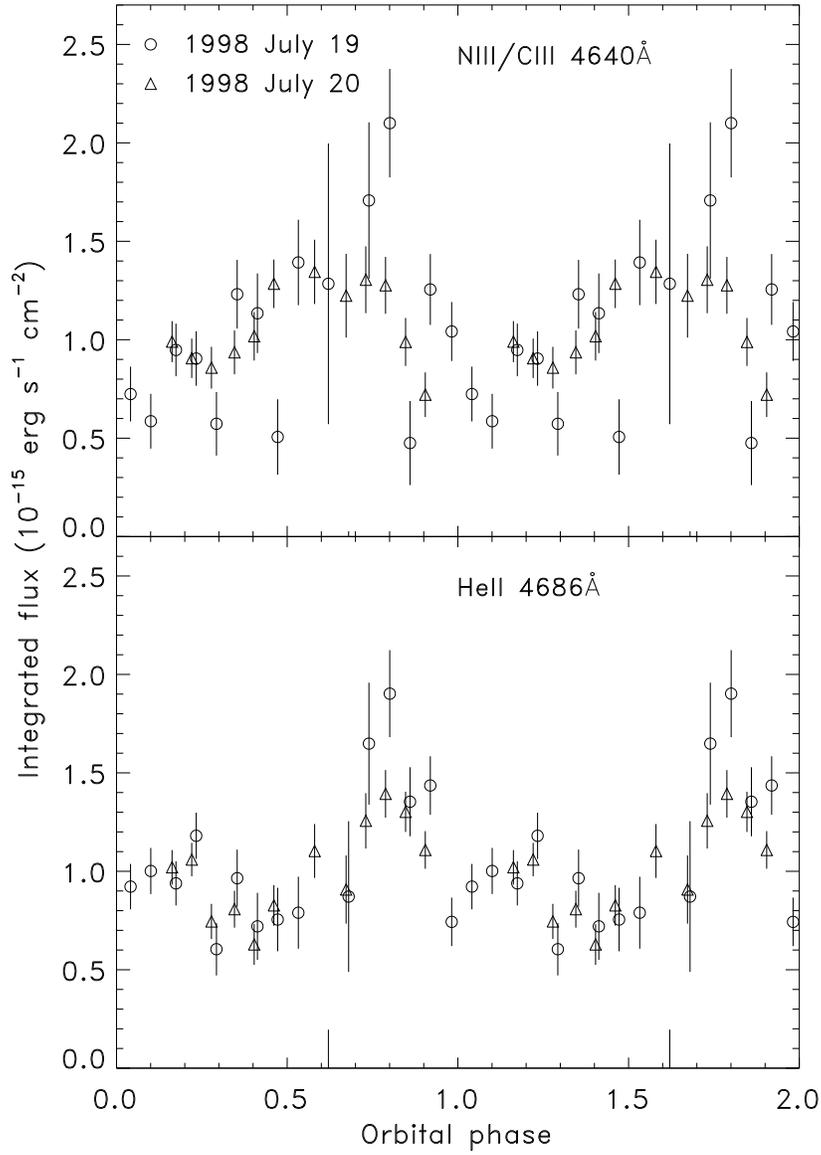,width=12cm}
\caption{Emission line light curves of N\,\textsc{iii}/C\,\textsc{iii}
and He\,\textsc{ii}.}
\label{LineLCFig}
\end{figure}

The light curves of the two lines shown in Fig.\ \ref{LineLCFig}
appear somewhat different in structure; the Bowen blend peak is
broader and earlier than that of He\,\textsc{ii}.  This suggests an
origin from different sites within the system.  The Bowen blend light
curve in fact appears similar to the continuum light curves shown in Fig.\
\ref{LCFig}; Bowen emission may therefore originate on the heated
face of the companion star, with the modulation arising from the
varying visibility of the heated region.  

He\,\textsc{ii} emission, shows 
a strong peak near phase 0.75 with a suggestion of a weaker one near
0.25.  The modulation probably indicates that the emission region is
optically thick.  We should be cautious in interpreting the light
curve, however, as the complex line behavior indicates multiple
emission sites.  The integrated light curve is an average of different
light curves of several regions.  A possible resolution to this
problem will be to use Doppler tomography to locate the dominant
emission sites (perhaps the two brightest spots).  Then we can fit a
toy model in which each of these spots (treated as a point source) is
allowed to vary smoothly in brightness over an orbital cycle.  This
procedure may allow us to approximately allow us to deconvolve the
light curves of the different regions.

\section{Doppler tomography}

\begin{figure}
\centering
\psfig{file=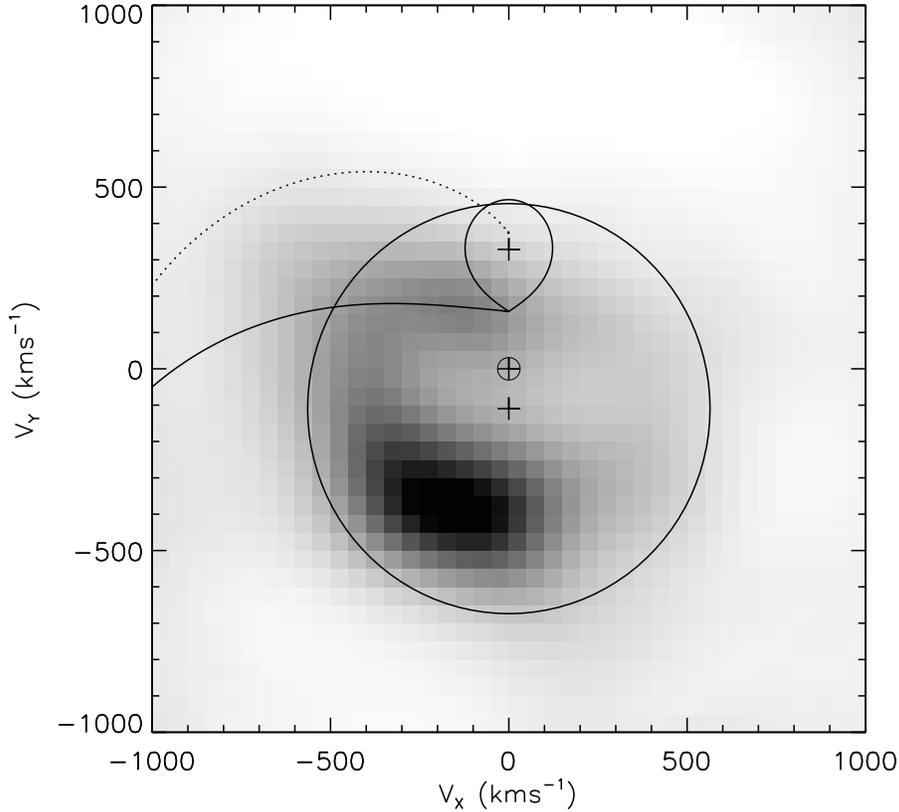,width=12cm}
\caption{Maximum entropy He\,\textsc{ii} 4686\,\AA\ tomogram.
Overplotted are the center-of-mass ($\oplus$), the centers of the two
stars ($+$), the companion star Roche lobe, the ballistic accretion
stream velocity (solid), the Keplerian velocity along the stream
position (dotted) and the Keplerian velocity at the disk edge
(circle).  All of these are based on the parameters of Paper I:
$M_1=1.8\,M_{\odot}$, $M_2=0.6\,M_{\odot}$, $T=0.24821$\,d,
$i=74^{\circ}$, $r_{\rm disk}=0.73\,r_{\rm L1}$.}
\label{TomogramFig}
\end{figure}

We have used the technique of Doppler Tomography (Marsh \& Horne 1988,
\cite{MH88}) to identify emission sites in velocity space.  Both the
back-projection method implemented in \textsc{molly} and
maximum-entropy method of \textsc{doppler} give similar results, as
does the alternative maximum entropy implementation of Spruit
\cite{S98}.  Fig.\ \ref{TomogramFig} shows a maximum entropy tomogram
generated with \textsc{doppler}.  Appropriate values were chosen for
instrumental resolution ($\sim180$\,km\,s$^{-1}$) and phase smearing.
We overplot the position of the Roche lobe of the companion, the
accretion stream ballistic velocity, the Keplerian velocity along the
accretion stream and the Keplerian velocity around the disk
edge. These are derived from uncertain system parameters, so they
should be viewed cautiously.  The parameters are determined from
light curve fits (see Paper I).  We should also beware that one of the
fundamental assumptions of Doppler tomography, that we always see all
of the line flux at {\it some} velocity, is clearly violated, as the
integrated line flux is not constant.

In spite of these cautions, we can learn something from the exercise.
The dominant emission site (corresponding to the main S-wave) appears
on the opposite side of the neutron star from the companion.  It is
inconsistent with the heated face of the companion and the stream/disk
impact point, although a tail does appear to extend upwards towards
the expected stream position.  As the emission appears to form an arc
roughly centered on the neutron star position, it is tempting to
associate it with asymmetric disk emission.  Unfortunately the
velocity of the strongest emission is too low for disk material.  This
can be seen from the fact that it lies inside the circle representing
the Keplerian velocity at the disk edge; the inner disk will have {\em
higher} velocities than this.  If the observed bright spot is indeed
emission from the disk then it must come from sub-Keplerian material.
A more promising explanation is suggested by the similarity to some SW
Sex type cataclysmic variables (e.g.\ V1315 Aql; compare H$\beta$
tomograms in Dhillon, Marsh and Jones 1991, \cite{DMJ91}, and Hellier
1996, \cite{H96}).  This is that the emission is actually associated with an
extension of the accretion stream beyond its nominal disk impact
point.  One possible model involves a disk-anchored magnetic
propeller (Horne 1999 \cite{H99}), suggested for SW Sex systems, which
ejects some of the stream material from the system.  An alternative is
that some material splashes from the stream impact point, rising high
above the disk.  Such material will follow a trajectory similar to
that seen, with the brightest observed spot corresponding to the point
where this splashing material reimpacts the disk.

\section{The spectral energy distribution}

\begin{figure}
\centering
\psfig{file=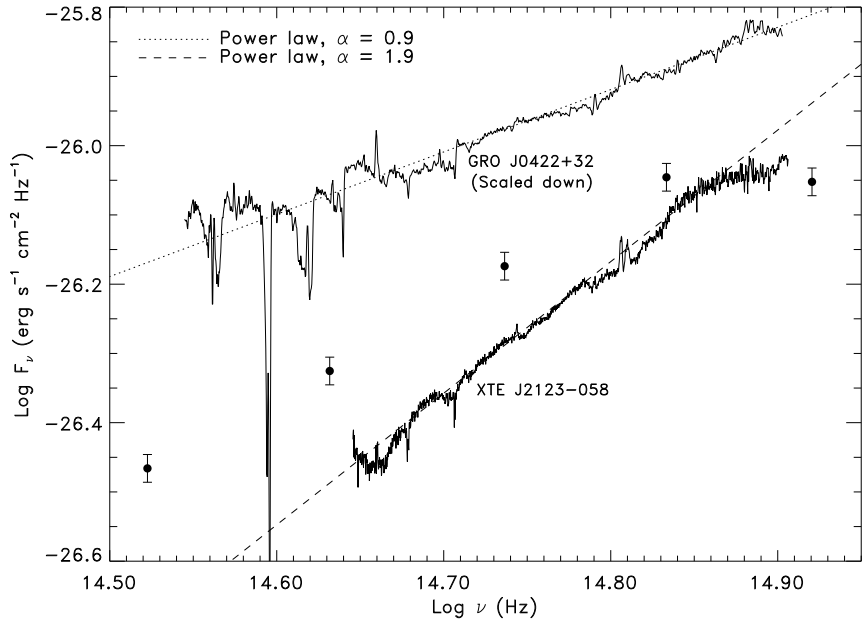,angle=90,width=12cm}
\caption{Dereddened spectral energy distributions of XTE~J2123--058
and the black-hole X-ray transient GRO~J0422+32.  Also plotted are the
photometry of Tomsick et al.\ (1998a, \cite{T98a}) and approximately
fitting power-law spectra for comparison.}
\label{SEDFig}
\end{figure}

We show in Fig.\ \ref{SEDFig} our average spectrum after dereddening
using the extinction curve of Cardelli et al.\ 1989, \cite{CCM89}) and
our reddening estimate of $\rm E(B-V)=0.12$ derived from the Na\,D1
line (see above).  We also show the photometry of Tomsick et al.\
(1998a, \cite{T98a}) from 1998 June 30 and a spectrum of GRO~J0422+32
(1992 August 17) provided by C.R. Shrader.  This black hole X-ray
transient has a 5.1-h orbital period, making it the black hole system
most similar to XTE~J2123--058.  Both the photometry
of Tomsick et al. and our spectroscopy appear steeper than the
spectrum of GRO~J0422+32, taking a steep blue power-law form.  In view
of the difficulties of accurately calibrating U band data, it is
unclear whether the apparent flattening off of the spectrum at high
energies is real or an artifact.

\section{Continuum behavior}

\begin{figure}
\centering
\psfig{file=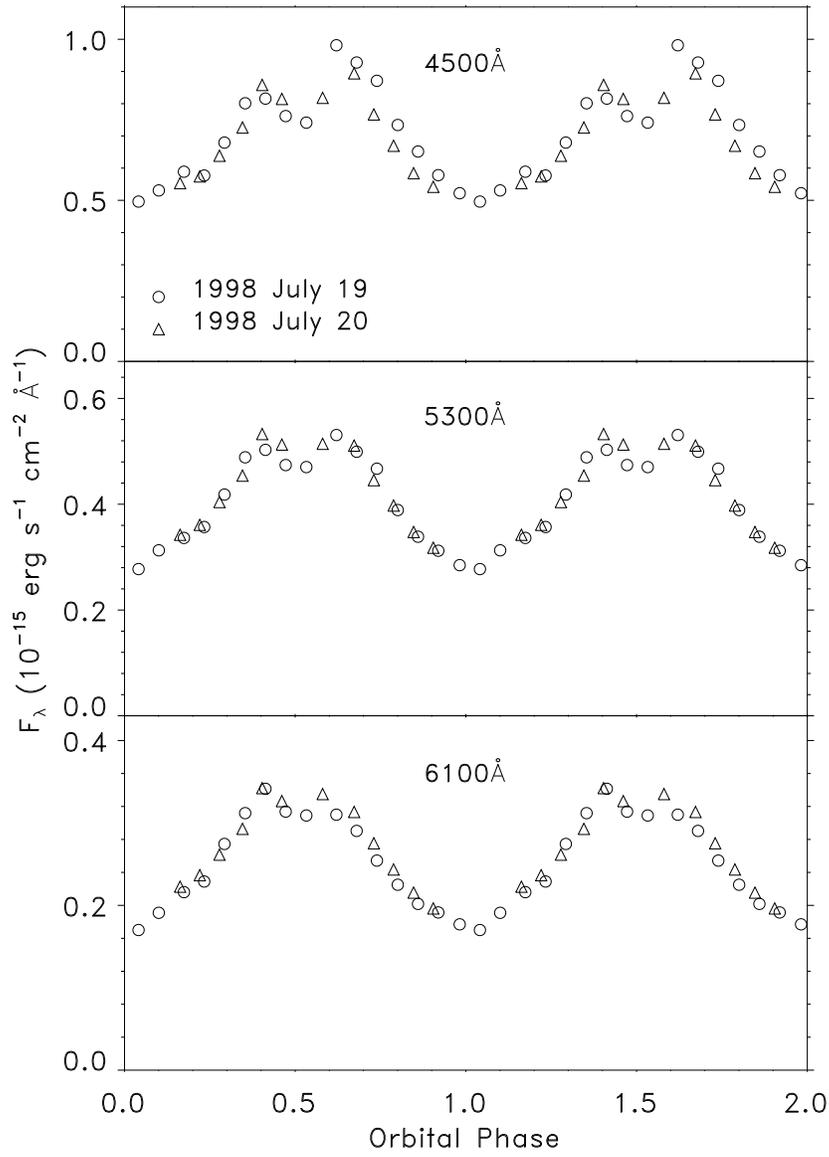,width=12cm}
\caption{Continuum light curves.  Formal errors are smaller than the
points, but systematic errors are larger, and are probably responsible
for the apparent differences between wavelengths.}
\label{LCFig}
\end{figure}

We show in Fig.\ \ref{LCFig} light curves for three `continuum' bins
at 4500\,\AA, 5300\,\AA\ and 6100\,\AA.  The light curves show very
similar shapes, with no significant differences in profile or
amplitude within this wavelength range.  The apparent differences
between them, most noticeable around phase 0.6, are likely due to
calibration uncertainties: coverage on each night was approximately
from phase 0.6 through 1.6.

The light curve morphology is fully discussed in Paper I.  We believe
it is mainly due to the changing aspect of the heated companion star,
which is the dominant light source at maximum light, near phase 0.5.
At minimum light (phase 0.0) the heated face is obscured and we see
the accretion disk only.  At all phases the unilluminated parts of the
companion star are expected to contribute negligible flux as the
outburst amplitude is $\sim 5$ magnitudes.

The lack of strong color dependence then indicates that either the
disk and heated companion have similar temperatures, or that both are
sufficiently hot that we only see the $F_{\nu} \propto \nu^{2}$
Rayleigh-Jeans part of the spectrum.  The very steep spectral energy
distribution shown in Fig.\ \ref{SEDFig} suggests that the latter is
the case.

\section{Conclusion}

Our main findings are summarized below.  Where appropriate, we make
comparisons with the black hole X-ray transient, GRO~J0422+32.  This
system has a 5.1-h orbital period and may be the black hole system
most similar to XTE~J2123--058.  We also note some similarities to the
neutron star LMXB 4U~2129+47 which has a 5.2-hr orbital period and
shows similar large amplitude photometric variations to XTE~J2123--058.

\begin{itemize}

\item
High excitation emission lines dominate the spectrum (He\,\textsc{ii},
N\,\textsc{iii}/C\,\textsc{iii}, C\,\textsc{iv} (see upper left
panel).  No He\,\textsc{i} emission is seen.  The blue line spectrum
looks rather similar to that of 4U~2129+47 (Thorstensen \& Charles
1982, \cite{TC82}).

\item
He\,\textsc{ii} emission is dominated by a region that coincides with
neither the heated companion star, the ballistic accretion stream nor
a Keplerian disk.  This is unlike GRO~J0422+32 (Casares et al.\ 1995,
\cite{C95}), the only other transient for which outburst Doppler
tomography has been performed.  In GRO~J0422+32, He\,\textsc{ii}
emission appears to originate from the accretion stream/disk impact
point.  Our observations may possibly be explained by extension of the
stream beyond the initial impact point.  This may be similar to the
He\,\textsc{ii} emission in 4U~2129+47 (Thorstensen \& Charles 1982,
\cite{TC82}) which has a radial velocity modulation with similar phase
and amplitude to that we see.

\item
The continuum spectral energy distribution is very blue, and steeper
than in black hole X-ray transients such as GRO~J0422+32 (see Fig.\
\ref{SEDFig}).  This implies that both the heated face of the
companion star, and the disk, are hotter than in similar short-period
black hole systems.  Such a conclusion is supported by the dominance
of high-excitation emission lines.  We suggest tentatively that this
may indicate more efficient X-ray irradiation in this (neutron star)
transient compared to black hole systems.  This has previously been
suggested by King, Kolb and Szuszkiewicz (1997, \cite{KKS97}) as an
explanation for why most transient LMXBs contain black holes whereas
apparently all persistent sources contain neutron stars.  Before we
can claim this is a firm conclusion we will need to assess the
uncertainty in the spectral energy distribution and perform a more
systematic comparison with other systems taking into account
differences in X-ray luminosity.

\end{itemize}

\section*{Acknowledgments}

Thanks to Chris Shrader for providing the spectrum of GRO J0422+32 for
comparison.  Doppler tomography used \textsc{molly} and
\textsc{doppler} software by Tom Marsh and \textsc{dopmap} software by
H.C. Spruit (\cite{S98}).  The William Herschel Telescope is operated
on the island of La Palma by the Isaac Newton Group in the Spanish
Observatorio del Roque de los Muchachos of the Instituto de
Astrofisica de Canarias.  RIH is supported by a PPARC Research
Studentship. 

\section*{References}

%%% Standard "thebibliography" LaTeX environment. References sorted BY
%%% ORDER OF APPEARANCE for hypertext links and cross-referencing.
%%%

\end{document}